# Is the Aether Entrained by the Motion of Celestial Bodies? What do the Experiments Tell Us?[1]

## Joseph Levy

4 square Anatole France, 91250 St Germain lès Corbeil, France
Email: levy.joseph@orange.fr

Even though the concept has evolved and if the designation as aether is improperly regarded as outdated, nobody today considers that the vacuum is empty. However, the nature and the properties of the substratum, which permeates the entire universe, remain for the most part unspecified. For example, divergent opinions are put forward by physicists about a possible dragging of the aether by the translational motion of celestial bodies due to gravitation. We show in this text that such a hypothesis is inconsistent with well established experimental data which, on the contrary, lend support to non-entrained aether theory based on Lorentz contraction. A revaluation of the aether drift to which the Earth is subjected is carried out.

## I. Introduction

Since the publication of Einstein's basic article "On the electrodynamics of moving bodies" in 1905, the aether has been excluded from the area of physics, being regarded as inexistent or at least inactive. Such an attitude signified that the laws of physics could be formulated in the same way, that the aether exists or not, a point of view expressed by Einstein in the following terms:

"The introduction of the luminiferous aether will prove to be superfluous inasmuch as the view here to be developed will not require an absolute space provided with special properties, nor assign a velocity-vector to a point of the empty space in which electromagnetic processes take place" [1].

This approach appeared quite revolutionary in 1905, since it called into question the ideas developed by a number of classical physicists such as Hooke, Lavoisier, Young, Huygens, Laplace, Fresnel, and Lorentz among others.
Certainly, the existence of a medium such as the aether, exhibits features difficult to explain since it was established, following Fresnel, that the Electromagnetic waves are transverse waves, whereas it is generally admitted that transverse waves need a solid medium to be transmitted. Yet, to regard the aether as a solid medium appeared to most physicists an idea quite difficult to accept.
However, although being a small minority, a fringe of contemporary physicists, never abandoned their belief in the aether; among them Ives [2], Janossy [3], Whittaker [4], Builder [5] Prokhovnik [6], Beckmann [7].
Dirac himself, in a controversy with Infeld in 1951, and in a paper published in Nature, strongly supported the hypothesis of the aether [8].

---

[1] With respect to the previous versions, further explanations are given and a misprint page 17 has been corrected.

In fact, despite its properties that seem so different from ordinary matter, a number of arguments speak in favour of a substratum [9] and these arguments have multiplied in the early twentieth century with the development of quantum mechanics. It is difficult, indeed, to accept that a "vacuum", endowed with physical properties such as permittivity and permeability may be empty. The ability of such an empty vacuum to transmit electromagnetic waves is also doubtful.

Quantum mechanics, on its part, regards the vacuum as an ocean of pairs of fluctuating virtual particles-antiparticles of very small life-time, appearing and disappearing spontaneously, which can be interpreted as a gushing of the aether, although the aether is not officially recognized by quantum mechanics. The interaction of the electrons and the vacuum, in particular, is regarded as the cause of the shifting of the alpha ray of the hydrogen atom spectrum, referred to as lamb shift [10]. The fluctuations of the vacuum are also assumed to explain the Casimir effect [11], and the Davies Fulling, Unruh, effect [12].

Einstein himself around 1916 changed his mind as regards the hypothesis of the aether. In a lecture given at the University of Leiden he declared [13]:

According to the general theory of relativity, space without aether is unthinkable for, in such space, there not only would be no propagation of light, but also no possibility of existence for standards of space and time (measuring rods and clocks).

A proof of the undeniable existence of the aether was given in ref [14]. Thus, the question to be answered today is not to verify its existence, but rather to specify its nature and its properties, and, in the first place, to determine if it is entrained (or not) by the translational motion of celestial bodies due to gravitation.

A number of arguments have been raised against entrained aether theory at the end of the 19° century and at the beginning of the twentieth century. Among the opponents of this theory were Larmor [15], Lorentz [16], Poincaré [17] and Lodge [18]. According to Lorentz, no aberration of starlight could occur if the aether was entrained. Lodge, in turn, challenged the theory on the basis of experimental evidence, showing that the speed of light was not changed in the vicinity of a rotating wheel.

Arguments in favour of non-entrained aether theory (NEAT) were also highlighted especially in the second half of the twentieth century by Builder [5] and Prokhovnik [6], who demonstrated that the existence of a fundamental aether frame and an aether drift, together with Fitzgerald-Lorentz contraction, imply that the two-way transit time of light in vacuum along a rod is independent of the orientation of the rod, in agreement with what Michelson-Morley experiments performed with highly rarefied gases verify.

Conversely, given the assumption of a fundamental aether frame and an aether drift, the constant value of this round-trip time along a rod in vacuum, whatever the orientation of the rod, implies necessarily Lorentz contraction. Indeed, let us suppose that we don't know a priori if the length of the rod varies (or not) when its orientation changes; as we know, according to NEAT, the value of the round-trip light time in a direction perpendicular to the Earth absolute motion is $\frac{2\ell}{C\sqrt{1-V^2/C^2}}$ while in the parallel direction, it is $\frac{2\Lambda}{C(1-V^2/C^2)}$. The equality of these two times requires necessarily that $\Lambda = \ell\sqrt{1-V^2/C^2}$.

However, for the supporters of entrained aether theory, none of the arguments put forward by its opponents were regarded as indisputable.

In 1845, Stokes proposed an approach assuming an incompressible and irrotational aether [19] which was considered later compatible with the experiment of Lodge, being entrained by the translation of the Earth but not by its rotation. Des Coudres and Wien [19] in 1898 assumed that aether dragging is proportional to the gravitational mass.

An objection to Stokes theory has been raised by different authors, among others by Lorentz [16B] who declared: "the irrotational motion of an incompressible fluid is completely determinate when the normal component of the velocity at its boundary is given: so that, if the aether were supposed to have the same normal component of velocity as the Earth, it would not have the same tangential component of velocity. It follows that no motion will in general exist which satisfies the Stokes' conditions".

Max Planck tried to rescue entrained aether theory [20], arguing, in a letter to Lorentz (1899), that the irrotational aether might not be incompressible but condensed by gravitation in the vicinity of the Earth. According to him, this could respond to the objections raised by Lorentz against the theory of Stokes. This new approach gave rise to the Stokes-Planck aether theory which based its views on the argument that it was consistent with Lodge's experiment and with the experiments of Fizeau and Sagnac, because the motion of objects of small mass was assumed not to drag the aether, contrary to the massive celestial bodies where they were placed.

In his response to Max Planck, Lorentz declared [21]: "If we wish to maintain the theory of Prof. Stokes by the supposition of a condensation in the neighbourhood of the Earth, it will be necessary to add a second hypothesis, namely that the velocity of light is the same in the not condensed and in the condensed aether. This is the theory that may be opposed to that of Fresnel, according to which the aether has no motion at all"…Lorentz also pointed out that "If we hope some time to account for the force of gravitation by means of actions going on in the aether, it is natural to suppose that the aether itself is not subject to this force. On these and on other grounds, I consider Fresnel's theory as the most satisfactory of the two. Prof Planck is of the same opinion".
But the supporters of entrained aether theory, following Petr Beckmann [7], considered that none of the above objections could prove definitively that the aether is not entrained. Certainly these authors, either did not know the experiments of Smoot and co-workers who highlighted an aether drift blowing at 370 Km/sec (namely 1,335,600 Km/h) at 20 Km altitude, or had not analysed in detail their consequences. Most of them believed that the aether is subject to gravity, so that it should not be affected by the demonstration of a distant aether drift that the experiments of Smoot highlighted.
However, as we shall see, the idea of an aether subject to gravity does not stand up to analysis. Moreover, even if one gave credit to such a concept, the front of the Earth would be faced to the pressure exerted by the non-entrained aether, which, in all likelihood, would thwart the entrainment. (see appendix 2).

The main goal we shall pursue in this text is to show that decisive arguments, susceptible of refuting the hypothesis of an aether entrained by the effect of gravitation, can be developed on the basis of a thorough analysis of Michelson-Morley experiments performed in air and in highly rarefied gases.
As we saw, modern aether drag theories, following the Stokes-Planck approach, consider that the aether is entrained by the Earth translation but not by the Earth rotation: the rotational velocity being less than 0.5 km/sec is assumed to give rise to an aether drift of this order of magnitude. Although this velocity is quite small relative to the speed of light, it can be several hundred km/h at certain latitudes and the research results for a possible detection can provide useful information. Conversely, Miller's approach reports an aether drift of greater magnitude. The relevance of these two approaches will be examined successively.

It is important here to specify that our study takes into account Hoek's law, which is the expression of Fresnel's law when measurements are carried out in the Earth frame [22], and it does not ignore the role of the refractive coefficient which, although close to 1 in gases, is heavily involved in the search for an exact determination of the aether drift [23]. Besides, second order terms which were ignored by Hoek and are required by the issue addressed, will be accounted for.

## II. Comparison of Michelson-Morley experiments operating in vacuum and in gas-mode according to NEAT and entrained aether theory

(Note that "vacuum" is a time-honoured expression which means the absence of gas or air, but does not imply the absence of aether).

Numerous very precise experiments in highly rarefied gases performed in the last decades, point out, with increasing accuracy, that no fringe shift should be observed in a perfect vacuum [24]. In contrast, Michelson-Morley experiments in gas mode [25-28] (air or helium) always gave positive results. The latter were disregarded by the classical authors, because the aether drift they deduced from them was small in comparison with that which was expected, so that they were often treated as experimental errors. But in their analysis they had neglected to take account of length contraction and of the refractive coefficient of the gases used.

More recently, some authors had the merit of revealing that the differences existing between vacuum and gas-mode experiments could be highlighted only if these factors were taken into account [23]. These authors conducted their analysis on the basis of non-entrained aether theory, pointing out that, in gas-mode, Lorentz contraction should not exactly conceal the presence of the aether drift, so that the latter could be measured. The results they obtained, taking account of length contraction and of the refractive coefficient of air (or alternatively of other gases), differed from the classical measurements, being of a greater magnitude.

In the following chapters, we shall check the pertinence of these statements, on the basis of a reanalysis of the methods used to determine the Earth absolute velocity. On the other hand, we shall test whether entrained aether theories can account for the significant differences observed between experiments performed with gases at normal pressure and with highly rarefied gases.

Overview of the procedure

Let us briefly bear in mind the modus operandi of Michelson's interferometer (Fig 1, 2 and 3). A beam of monochromatic light emanated from a source is split into two beams travelling in perpendicular directions by a beam splitter. The first beam passes through the beam splitter, before reflecting in the mirror M, and after a second reflection in the beam splitter, heads toward the telescope. The second beam reflects first in the beam splitter and then in the mirror N, before travelling toward the telescope. In the telescope the beams may interfere giving rise to an interference pattern.

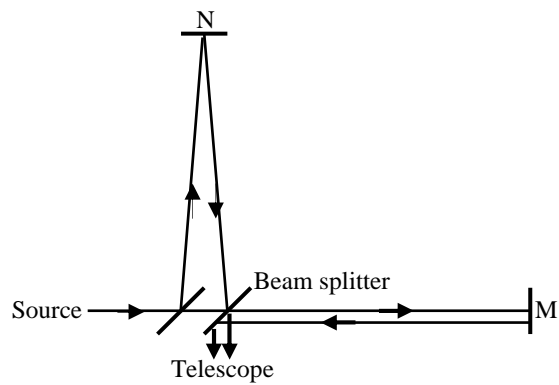

Fig 1. Transit of the light beams relative to the aether in the two arms of a Michelson-Morley interferometer. (This schematic diagram corresponds to the first position of the interferometer shown in figure 2).

The experiment involves two steps: in the first step the longitudinal arm OM is aligned along the Earth absolute velocity vector (Figures 1 and 2), (see footnote 1); in the second step the interferometer is subjected to a 90° rotation: the arm OM is now aligned perpendicularly to the direction of the Earth velocity vector, while it is the arm ON which is now parallel to it (Figure 3). We assume that the arms are constructed to have equal length when they are parallel to each other. Yet, insofar as Lorentz contraction applies, a small difference will occur when they are placed in a position perpendicular to one another.

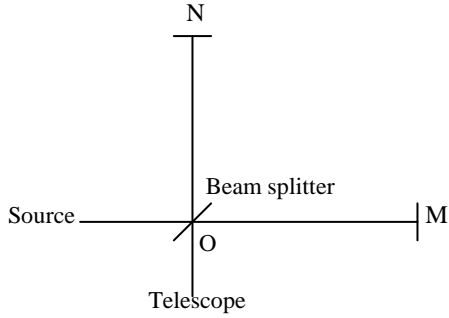 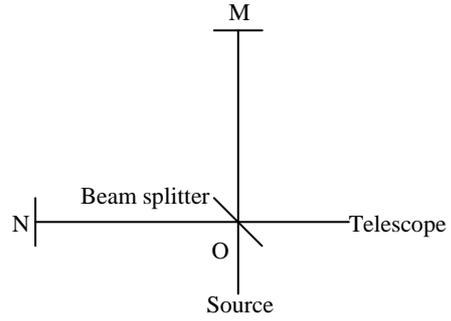

Fig 2. A. First position of the interferometer   Fig 3. B. Second position of the interferometer

(Note that, in what follows, the subscript 1 will be restricted to the light transit time along the arm OM, and the subscript 2 to the arm ON, regardless of their position and of their length).

### II. 1. Analysis of Michelson-Morley experiments operating in vacuum and in gas-mode according to entrained aether theory

Before pursuing, recall that the appellation "entrained aether theory" indicates that the aether is supposed to be globally entrained by the Earth translational motion. In contrast, the idea of an entrainment by the Earth rotation was abandoned, having been refuted by Lodge's experiment which proved that the speed of light is not modified in the near proximity of a rotating wheel.
These assumptions, which were those of Stokes-Planck aether theory, imply that a small aether drift, equal to the rotational speed of the Earth, at the place where the experiment is performed, should exist. In the text which follows, we will check whether this requirement is compatible with other well-established experimental data.
Recall also that Stokes-Planck aether theory does not recognize Lorentz contraction.

### II. 1. 1. Vacuum experiments

Assuming first that the arm OM is aligned in the direction of the Earth rotation, and denoting by $T_1$ the two-way transit time of light along this arm, we have (Figures 1 and 2):

$$T_1 = \frac{L}{C-v} + \frac{L}{C+v} = \frac{2L}{C(1-v^2/C^2)}, \qquad (1)$$

where $L$ is the length of the arm and $v$ the value of the Earth rotational velocity at the location of the experiment, (which is generally between 0.2 and 0.47 km/sec). $C - v$ and $C + v$ are the values of the speed of light respectively in the direction of rotation and in the opposite direction.

Two-way transit time of light along the arm ON:
While the Earth surface has covered the distance $d = vt_2$ relative to the irrotational aether, the light signal has covered in this aether the oblique path $D = Ct_2$ from the beam splitter to the end of the arm ON (Figures 1 and 2). Pythagoras theorem then yields:

$$C^2 t_2^2 - v^2 t_2^2 = L^2,$$

So, the two-way transit time of light $T_2 = 2t_2$ in this direction yields:

$$T_2 = \frac{2L}{C\sqrt{1-v^2/C^2}}. \qquad (2)$$

The difference between the round-trip light times in the two orthogonal directions is:

$$\delta T = \frac{2L}{C(1-v^2/C^2)} - \frac{2L}{C\sqrt{1-v^2/C^2}}.$$

Ignoring the terms multiplying $2L/C$ by the 4° order terms in $v/C$ this expression yields:

$$\delta T \approx \frac{2L}{C}[(1+\frac{v^2}{C^2}) - (1+\frac{1}{2}\frac{v^2}{C^2})].$$

$$\approx \frac{Lv^2}{C^3}. \qquad (3)$$

Upon the 90° rotation of the interferometer (Figure 3), the displacement of the interference fringes is multiplied by two, giving:

$$\Delta T \approx 2\frac{Lv^2}{C^3} \qquad (4)$$

This result shows that, since Stokes-Planck aether theory assumes no Lorentz contraction, and therefore is not affected by the masking effect that results, the aether drift predicted by this theory, should give rise to a fringe shift in vacuum experiments, providing a means to test its validity.

As we saw, this theory maintains that the magnitude of the aether drift should be equal to the Earth rotational velocity at the location of the experiment, which is about 0.2 to 0.47 km/sec at the usual locations, or, more precisely, 720 to 1674 km/h.

The corresponding value of the round-trip light speed anisotropy is:

$$\Delta C = C[\sqrt{1-v^2/C^2} - (1-\frac{v^2}{C^2})]$$

Ignoring the terms of fourth order, this expression yields:

$$\Delta C \approx C[(1-\frac{1}{2}\frac{v^2}{C^2}) - (1-\frac{v^2}{C^2})] \approx \frac{v^2}{2C}$$

So that

$$\frac{\Delta C}{C} \approx \frac{v^2}{2C^2}$$

For a value of the rotational Earth velocity equal to 0.47 km/sec, the ratio $\Delta C/C$ predicted by Stokes-Planck entrained aether theory should be equal to $1.22 \times 10^{-12}$, which corresponds to an anisotropy $\Delta C \approx 3.66 \times 10^{-7} km/\sec$. This is the value that would be observed in vacuum if the theory was consistent.

However, modern Michelson-Morley experiments in highly rarefied gases set a limit on the ratio $\Delta C/C$ equal to $1 \times 10^{-17}$ [24], which corresponds to an anisotropy $\Delta C \approx 3 \times 10^{-12} km/\sec$, less than $10^5$ times smaller.

Such a tiny anisotropy in rarefied gases can be interpreted as perfect isotropy in vacuum, and, given that no Lorentz contraction is assumed by Stokes-Planck aether theory, it means absence of aether drift and therefore total entrainment of the aether, contrary to what Lodge's experiment requires.

Therefore Stokes-Planck aether theory turns out to be inconsistent with Michelson-Morley experiments performed in highly rarefied gases.

In contrast, as we shall see, only a theory based on Lorentz contraction can explain that perfect isotropy in vacuum does not imply absence of aether drift, allowing to reconcile Lodge's experiment with modern Michelson-Morley experiments performed in vacuum.

This result, which proves the necessity of Lorentz contraction -a physical process closely associated with non-entrained aether theory- argues in favour of NEAT, and refutes Stokes-Planck aether theory which does not assume it.

This is a first objection against entrained aether theory.
A second objection will be raised below.

**II. 1. 2. Gas-mode experiments**

In order to develop our second argument (by contradiction) we will again suppose a priori that the small aether drift caused by the Earth rotation exists, and test whether this assumption is pertinent.

Besides the conventional assumptions of Stokes-Planck aether theory, the role of the refractive coefficient will be taken into account.
To first order, Hoek's experiment showed that the speed of light in refractive media at rest on Earth but moving relative to the aether at speed $v$, is $\dfrac{C}{n} - \dfrac{v}{n^2}$ in the direction of the Earth absolute motion, and $\dfrac{C}{n} + \dfrac{v}{n^2}$ in the opposite direction [14, 22], (see chapter **IV**). In Stokes-Planck entrained aether theory, the direction of the Earth absolute motion is replaced by the direction of the Earth rotation.
Here, the investigation is designed to detect and specify second order effects. So, a modified Hoek's formula which takes account of those effects is required. (See appendix 1).

In the first step, the arm OM is aligned in the direction of the Earth rotation (Figures 1 and 2).
With the hypothesis, of the small aether drift due to the rotation of the Earth, the modified Hoek's expression for the velocity of light in air (or in gas) is $\dfrac{C}{n}\sqrt{1 - \dfrac{v^2}{C^2}(1 - 1/n^2)} - \dfrac{v}{n^2}$ in the direction of rotation, and $\dfrac{C}{n}\sqrt{1 - \dfrac{v^2}{C^2}(1 - 1/n^2)} + \dfrac{v}{n^2}$ in the opposite direction, where $v$ designates the rotational speed of the Earth at the place of the experiment. (see chapter IV and the appendix 1).
Therefore, for entrained aether theory the–two-way transit time of light along this arm is equal to:

$$T_1' = \dfrac{L}{\dfrac{C}{n}\sqrt{1 - \dfrac{v^2}{C^2}(1 - 1/n^2)} - \dfrac{v}{n^2}} + \dfrac{L}{\dfrac{C}{n}\sqrt{1 - \dfrac{v^2}{C^2}(1 - 1/n^2)} + \dfrac{v}{n^2}}$$

Ignoring the terms multiplying $\dfrac{2L}{C/n}$ by the fourth order terms in $v/C$, this expression yields:

$$= \dfrac{2L\dfrac{C}{n}\sqrt{1 - \dfrac{v^2}{C^2}(1 - \dfrac{1}{n^2})}}{\dfrac{C^2}{n^2}[1 - \dfrac{v^2}{C^2}(1 - \dfrac{1}{n^2})] - \dfrac{v^2}{n^4}} \approx \dfrac{2L\dfrac{C}{n}[1 - \dfrac{1}{2}\dfrac{v^2}{C^2}(1 - \dfrac{1}{n^2})]}{\dfrac{C^2}{n^2}(1 - \dfrac{v^2}{C^2})},$$

$$T_1' \approx \dfrac{2L}{C/n}[1 + \dfrac{1}{2}\dfrac{v^2}{C^2} + \dfrac{1}{2}\dfrac{v^2}{n^2 C^2}]. \tag{5}$$

Two-way transit time of light along the arm ON.
Our analysis shows that due to the presence of air (or gas) moving at speed $v$ relative to the aether, the speed of light along the arm perpendicular to the Earth rotation is:

$$\dfrac{C}{n}\sqrt{1 - v^2/C^2},$$

(see appendix 1).
So that, the two-way transit time of light along the arm ON is:

$$T_2' = 2t_2 = \frac{2L}{\frac{C}{n}\sqrt{1-v^2/C^2}} \approx \frac{2L}{C/n}(1+\frac{1}{2}v^2/C^2). \qquad (6)$$

The round-trip light time difference between the two orthogonal directions is:

$$\delta T' \approx \frac{2L}{C/n}[(1+\frac{1}{2}\frac{v^2}{C^2}+\frac{1}{2}\frac{v^2}{n^2C^2}) - (1+\frac{1}{2}\frac{v^2}{C^2})].$$

Thus:

$$\delta T' \approx \frac{Lv^2}{C^3}(\frac{1}{n}). \qquad (7)$$

Upon the 90° rotation of the interferometer (Figure 3), the displacement of the interference fringes is multiplied by two, giving:

$$\Delta T' \approx \frac{2Lv^2}{C^3}(\frac{1}{n}).$$

Given that the refractive coefficient of air is equal to 1.00028, its value, for air, yields:

$$\Delta T' \approx 0.9997(\frac{2Lv^2}{C^3}).$$

The fact that this expression is hardly different from $\Delta T$ (formula (4)) is at odds with the experiment for which, *a clear difference*, *as regards the fringe shifts*, is observed between measurements performed in air [25, 26, 28] and in highly rarefied gases [24, 28] (see chapter **II.2**). The near-identity of $\Delta T$ and $\Delta T'$ constitutes a second objection which permits to refute entrained aether theory. (Besides, this objection would persist even if the aether drift highlighted by Lodge was overestimated and was much closer to zero).

We shall now check whether NEAT can better account for the experimental results.

**II. 2. Analysis of Michelson-Morley experiments operating in vacuum and in gas-mode according to NEAT.**

For this analysis, we must first determine the direction of the Earth absolute velocity vector[2].

**II. 2. 1 Vacuum experiments**

In order to explain the quite small fringe shifts observed when interferometers are filled with highly rarefied gases, which augur a null fringe shift in a perfect vacuum, NEAT, following Lorentz, Fitzgerald and Larmor, assumes length contraction.

Provided that the arm OM is aligned along the Earth absolute velocity vector (Figures 1 and 2), the two-way transit time of light along this arm can be easily calculated, as follows[3]:

---

[2] In practice this research is fraught with difficulties of application, because it requires trial and error to find the direction by which the sought fringe shift is maximum. Furthermore the direction of the absolute velocity of the Earth may make a non-zero angle with the plane where the interferometer lies, which depends on the instant of the experiment. But, for our theoretical study, a perfect determination assuming that the angle is zero and that the longitudinal arm is aligned along the Earth absolute velocity vector can be put to the proof, and it is quite appropriate because it can objectify the maximum differences existing between NEAT and entrained aether theory.

[3] The independence of the round-trip light time relative to the orientation, which is expected in vacuum along a rod subject to Lorentz contraction, was demonstrated in ref [34]. For our purpose, only the direction of the Earth absolute velocity and the orthogonal direction are necessary.

$$T_1 = \frac{L\sqrt{1-V^2/C^2}}{C-V} + \frac{L\sqrt{1-V^2/C^2}}{C+V} = \frac{2L}{C\sqrt{1-V^2/C^2}}.$$

Here, contrary to paragraph II.1.2, *V* is the speed of the Earth frame relative to the fundamental aether frame.

Two-way transit time of light along the arm ON:
While the Earth has covered the distance $d=Vt_2$ relative to the fundamental aether frame, the light signal has covered in this frame the oblique path $D=Ct_2$ from the beam splitter to the end of the arm ON (Figures 1 and 2). Pythagoras theorem then yields:

$$C^2 t_2^2 - V^2 t_2^2 = L^2.$$

So, the two-way transit time of light $T_2 = 2t_2$ in this direction is:

$$T_2 = \frac{2L}{C\sqrt{1-V^2/C^2}}.$$

Thus, according to NEAT, the two-way transit times of light in vacuum along the two arms of a Michelson interferometer are identical and no fringe shift is expected, as the experimental investigation confirms.
Therefore:

$$\Delta T = T_1 - T_2 = 0. \tag{8}$$

(This result is not affected by erroneous measurements given that there is no direct measurement of lengths, and because clock retardation affects equally the two orthogonal paths).

Of course, as the demonstration shows, this does not mean that there is no aether drift in vacuum: the existence of the aether drift is not dependent on the presence of gas. Simply, by shortening the longitudinal arm, Lorentz contraction renders the two orthogonal travel times identical in vacuum, preventing to highlight the effect of the aether drift.

Note also that all length measurements which are made in the Earth frame along the longitudinal arm, in order to determine the two-way speed of light, use contracted standards that do not allow to highlight length contraction, so that the longitudinal arm in Michelson experiment is found to be *L*. On the other hand, given that the time measurements are subject to clock retardation, the apparent two-way transit time of light is found to be:

$$T_1\sqrt{1-V^2/C^2} = \frac{2L}{C}.$$

Therefore the two-way speed of light along this arm is found equal to:

$$\frac{2L}{2L/C} = C$$

Along the transversal arm, there is no measurement error as regards the length because the arm is not contracted, but clock retardation applies. Finally the experimental result is the same as in the longitudinal arm. Thus the apparent two-way speed of light is found to be C in both arms, and

$$\Delta C = 0$$

in agreement with what the experiments of Schiller and Herrmann indicate.
Therefore, the isotropy of the speed of light is only apparent, but it is what the experiment shows.

We shall now see whether, in the presence of refractive media, the exact equality of the light transit times along the two arms of Michelson interferometer is (or not) maintained.

**II. 2. 2. Gas-mode experiments**

As in section II.1.2.the modified Hoek's formula which takes into account second order effects will be used**.**

Provided that the arm OM is aligned along the Earth absolute velocity vector (Figures 1 and 2), the two-way transit time of light along this arm is:

$$T_1' = \frac{L\sqrt{1-V^2/C^2}}{\frac{C}{n}\sqrt{1-\frac{V^2}{C^2}(1-1/n^2)} - \frac{V}{n^2}} + \frac{L\sqrt{1-V^2/C^2}}{\frac{C}{n}\sqrt{1-\frac{V^2}{C^2}(1-1/n^2)} + \frac{V}{n^2}}$$

where $\frac{C}{n}\sqrt{1-\frac{V^2}{C^2}(1-1/n^2)} - \frac{V}{n^2}$ is the speed of light in air (or alternatively in the gas where the light propagates) in the direction of the Earth absolute velocity, and $\frac{C}{n}\sqrt{1-\frac{V^2}{C^2}(1-1/n^2)} + \frac{V}{n^2}$ the speed of light in the opposite direction (see chapter IV and the appendix 1).

Ignoring the terms multiplying $\frac{2L}{C/n}$ by the fourth order terms in $v/C$, this expression yields:

$$T_1' = \frac{2L\frac{C}{n}\sqrt{1-\frac{V^2}{C^2}}\sqrt{1-\frac{V^2}{C^2}(1-\frac{1}{n^2})}}{\frac{C^2}{n^2}[1-\frac{V^2}{C^2}(1-\frac{1}{n^2})] - \frac{V^2}{n^4}} \approx \frac{2L\frac{C}{n}(1-\frac{1}{2}\frac{V^2}{C^2})[1-\frac{1}{2}\frac{V^2}{C^2}(1-\frac{1}{n^2})]}{\frac{C^2}{n^2}(1-\frac{V^2}{C^2})}$$

$$\approx \frac{2L}{C/n}(1+\frac{1}{2}V^2/n^2C^2).$$

Two-way transit time of light along the arm ON.
Our analysis shows that, due to the presence of air (or gas) moving at speed $V$ relative to the aether, the speed of light along the arm perpendicular to the Earth absolute motion is:

$$\frac{C}{n}\sqrt{1-V^2/C^2},$$

(see appendix 1).

The two-way transit time of light in this direction is therefore:

$$T_2' = 2t_2 = \frac{2L}{\frac{C}{n}\sqrt{1-V^2/C^2}} \approx \frac{2L}{C/n}(1+\frac{1}{2}V^2/C^2).$$

The round-trip light time difference between the two orthogonal directions yields:

$$\delta T' \approx \frac{2L}{C/n}[(1+\frac{1}{2}\frac{V^2}{C^2}) - (1+\frac{1}{2}\frac{V^2}{n^2C^2})].$$

$$\approx L\frac{V^2}{C^3}(n-\frac{1}{n}) \qquad (9)$$

Upon the 90° rotation of the interferometer (Figure 3), the displacement of the interference fringes is multiplied by two, giving:

$$\Delta T' \approx 2L\frac{V^2}{C^3}(n-\frac{1}{n}). \qquad (10)$$

Given that the refractive coefficient of air is 1.00028, the expression (10) relative to air yields:

$$\Delta T' = 5.6\times 10^{-4}(2L\frac{V^2}{C^3}).$$

The expression of the Earth absolute velocity deduced from (10) is:

$$V = \sqrt{\frac{\Delta T' C^3}{2L} \times \frac{1}{(n - 1/n)}}.$$

Therefore, contrary to entrained aether theory, NEAT predicts a significant difference between vacuum and gas-mode experiments: unlike vacuum experiments, the presence of air (or gas) prevents length contraction to exactly conceal the effect of the aether drift, allowing it to be measured.

NEAT also reconciles Lodge's experiment with the isotropy of the round-trip light velocity in vacuum. Thus, in contrast to entrained aether theory, NEAT accounts for the available experimental data.

In the following chapters we shall study the impact of NEAT in classical experiments.

### III. Critical study of the interpretation made by Michelson-Morley and Miller of their experiments

#### III. 1. <u>Michelson-Morley experiment</u>

In their attempts to measure the Earth absolute velocity, Michelson and Morley used an interferometer whose effective length of the arms measured approximately 11 meters, giving a total light path going and returning of 22 meters [25].

If there were no length contraction, the rotation of the interferometer would not affect the arms, but, otherwise, the arms shorten when they lie along the Earth absolute velocity vector. But these considerations did not concern the calculations made by Michelson and Morley [25] who did not take account of length contraction. Note also that the refractive coefficient of air was taken by these authors equal to 1.

With these assumptions, the round-trip light time difference between the two orthogonal directions found by Michelson and Morley after rotation of the interferometer was:

$$\Delta T = 2 \frac{L v^2}{C^3},$$

which yielded:

$$v^2 = \frac{\Delta T C^3}{2L}. \tag{11}$$

The original estimation of the Earth absolute velocity by Michelson and Morley in 1887 varied slightly, the maximum being 8 Km/sec, the mean values being 6.22 Km/sec at noon and 6.80 Km/sec at 18h.

If Michelson and Morley's analysis was unquestionable, no significant difference of fringe shifts would be observed between experiments, whether performed in air at atmospheric pressure or in highly rarefied air, which is not the case, since, contrary to those carried out in air at atmospheric pressure, the latter are much smaller and can be unobservable.

Only when one takes account of length contraction and of the refractive coefficient, a significant difference of fringe shifts is predicted, between experiments carried out in air at atmospheric pressure and in highly rarefied air, in agreement with observation.

Taking account of Lorentz contraction and of the refractive coefficient, Michelson and Morley's formula (11) is converted into:

$$V^2 = \frac{\Delta T C^3}{2L} \times \frac{1}{(n - \frac{1}{n})},$$

from (10). *Which means that the proper fringe shift observed by Michelson and Morley corresponded to an estimation of the Earth absolute velocity of much greater magnitude.*

With these corrections, Michelson and Morley's measurements yielded for *V* a value of the order of 264 *km*/sec at noon and 289 *km*/sec at 18h. As we shall see, the results obtained by Miller with an improved device departed neatly from these values.

Note also that these results concerned the mean values of the projection of the Earth absolute velocity vector on the plane of the interferometer, rather than the velocity vector itself. This remark also applies to the results obtained by Miller below.

### III. 2. **Miller's experiment**

Miller used an interferometer which by a system of reflections allowed a light path going and returning of 64.06 meters [26, page 209].

Like Michelson and Morley, Miller did not take account of length contraction and of the refractive coefficient of air in his calculations, so that his estimation of the Earth absolute velocity yielded also:

$$v_M = \sqrt{\frac{\Delta T C^3}{2L}}.$$

The magnitude of $v_M$ varied slightly at the different seasons of the year, the minimum being equal to 9.3 Km/sec in February, and the maximum to 11.2 Km/sec in August [26, p 230].

Like for Michelson and Morley, if Miller's analysis was unquestionable, no significant difference of fringe shifts would be observed between experiments whether performed in air at atmospheric pressure or in highly rarefied air, which is not the case, since, unlike those carried out in air at atmospheric pressure, the latter are much smaller and can be unobservable. Only NEAT can account for such a difference which is observed experimentally.

With the corrections brought when one takes account of length contraction and of the refractive coefficient, Miller's formula is converted into:

$$V_M = \frac{v_M}{(n - \frac{1}{n})^{\frac{1}{2}}},$$

from (10).

For the estimation of the aether drift made by Miller at Mount Wilson in February which was $v_M = 9.3 Km/\sec$, we obtain after correction, given that the refractive coefficient of air is equal to 1.00028:

$$V_M = \frac{9.3}{(n - \frac{1}{n})^{\frac{1}{2}}} Km/\sec = \frac{9.3}{\sqrt{5.6 \times 10^{-2}}} Km/\sec,$$

which gives $V_M \approx 395 Km/\sec$.

For the estimation made by Miller in August which yielded 11.2 Km/sec, we obtain after correction:

$$V_M \approx 476 Km/\sec.$$

**Important note**

The hypothesis that no fringe shift can occur in vacuum experiments because after crossing the armoured walls of the interferometer, the aether velocity would decrease instantaneously from 9.3km/sec (33480 km/h) to zero, is quite unrealistic and is not corroborated by facts [28,29]. Besides, if this were the case, due to the absence of drift, no fringe shifts could also occur in the above mentioned armoured interferometer even if it was filled with air or with gas, but this is not what has been observed [28].

Therefore, if no fringe shift is observed in vacuum experiments, this can only be due to the fact that the refractive coefficient is equal to 1. If so, the difference of fringe shifts observed in gas at atmospheric pressure and in highly rarefied gas can be explained only by non-entrained aether theory, which rests on Lorentz contraction.

**Discussion**

The existence of an aether drift which was demonstrated by Miller is not disputable today. As shown by Munera [27], after correction of systematic errors it was also evidenced by all the experiments in gas mode performed to date using different techniques. Although the authors of these experiments interpreted the fringe shifts they observed as negligible or due to experimental errors, a fact that reflected the knowledge of their time, a thorough analysis of these experiments does not corroborate this point of view.

In his book [30], Maurice Allais recognized the fallacy of the statement that Michelson-Morley experiment gave a null result. About the claims of Shankland [31] who attributed the effects obtained by Miller to local disturbances and specifically to disturbances resulting from temperature, Maurice Allais replied [30, Page 412]: "In regard to the extensive and thorough investigation I have done of the study of Shankland, this conclusion appears quite unfounded and distorted (biaisée) given the great consistence that appears in the analyses of the observations made by Miller, that could not result from local disturbances…"

And also: "Miller was a very experienced experimenter, who has thoroughly analyzed the potential effects of temperature, under very different experimental conditions".

Page 585 of his book Maurice Allais declared that the interferometric observations carried out by Miller have been confirmed by the 3 series of experiments performed by him and by those performed by Maurice Esclangon.

As shown in ref [32] the estimation by James DeMeo of Shankland's work agrees with that of Maurice Allais.

Therefore, there is no reason to cast doubt to the fact that the physicists responsible for Michelson and Morley's interferometry have highlighted an aether drift. *This fact alone is of utmost importance for the development of physics.* However, as regards the magnitude of the observed anisotropy, as well as the direction of the Earth absolute velocity vector, they were subject to different sources of error. One of them, as we have seen, was the ignorance of length contraction and of the role played by the refractive coefficient, and another one was the imprecise knowledge of the temporal variation of the angle between the Earth absolute velocity vector and the plane in which the interferometer lies. This imprecise knowledge explains, at least partly, the disparity in the estimations of the direction of solar motion made by different authors, as shown in Munera's 2009 article [27], page 6. See also Smoot's estimations [33].

Researches carried out today permit to obtain, step by step, a more precise evaluation of the equatorial co-ordinates.

There remains some points to clarify, but they do not call into question the light speed anisotropy observed by Miller which, after a thorough reanalysis, permits to verify the non-dragging of the aether.

## IV. Conclusion

1. Our analysis shows that entrained aether theory cannot account for the differences existing in rarefied gas and in gas at atmospheric pressure, which are observed experimentally whatever the assumptions made. Moreover it is proved defective by modern Michelson-Morley experiments, and it is incompatible with the aether drift highlighted by Smoot's experiments, even if this drift has been observed at a distance form the Earth (see appendix 2). Thus, a dragging of the aether by gravity is excluded.

2. The estimations of the aether drift we made after analysing the results obtained by Michelson and Miller, differ somewhat from those of Cahill and Kitto who used the expression $\frac{C}{n} - V$ for the speed of light in refractive media measured from the Earth frame in the direction of the Earth absolute velocity, and $\frac{C}{n} + V$ in the opposite direction. They are also different from Consoli's estimations who used $\frac{C}{n} - V(1-\frac{1}{n^2})$ and $\frac{C}{n} + V(1-\frac{1}{n^2})$. However, according to NEAT, and in agreement with Fresnel's law, the latter are the values obtained when measurements are made from the aether frame. In fact, as demonstrated by Hoek's experiment, the correct values when measurements are made on Earth are, to first order, $\frac{C}{n} - \frac{V}{n^2}$, and $\frac{C}{n} + \frac{V}{n^2}$ [14, 22]. Besides, as we saw, insofar as second order developments are needed, the modified Hoek's formulas $\frac{C}{n}\sqrt{1-\frac{V^2}{C^2}(1-\frac{1}{n^2})} - \frac{V}{n^2}$ and $\frac{C}{n}\sqrt{1-\frac{V^2}{C^2}(1-\frac{1}{n^2})} + \frac{V}{n^2}$ must be used. (See note[4] and demonstration in the appendix 1).

Our analysis also shows that the speed of light along the transverse arm, when it is immersed in a refractive medium moving relative to the aether frame at speed *V*, but at rest with respect to the arm, also differs from that of our predecessors, being equal to $\frac{C}{n}\sqrt{1-\frac{V^2}{C^2}}$.

**Appendix 1**

**Speed of light at instant *t* in a refractive medium at rest relative to the Earth frame, in a direction making an angle $\theta$ with respect to the Earth absolute velocity vector**

In a previous article [14] it has been shown that, to first order, the speed of light in moving refractive media in directions making at instant *t* an angle $\theta$ with respect to the Earth absolute velocity is equal to $\frac{C}{n} - \frac{V\cos\theta}{n^2}$ when the projection of the velocity vector of the light signal on the Earth absolute velocity points in the same direction as the latter, and to $\frac{C}{n} + \frac{V\cos\theta}{n^2}$ when it points in the opposite direction.

Here, the conditions of the problem under consideration require more accuracy. To this end, we shall search for terms of the form $\frac{C}{n} - \frac{V\cos\theta}{n^2} + \varepsilon$, (12), and $\frac{C}{n} + \frac{V\cos\theta}{n^2} + \varepsilon$, (13), and, after determination of the unknown $\varepsilon$, we shall verify if the expressions (12) and (13) comply with the requirements of the issue.

We start from the fact that in Hoek's experiment, there is no fringe shift in the interference pattern whatever the direction of the device.

According to NEAT, the second order expressions of the one-way speed of light in vacuum are:
$$\sqrt{C^2 - V^2\sin^2\theta} - V\cos\theta \qquad (14)$$
when the projection of the velocity vector of the light signal on the Earth absolute velocity, points in the same direction as the latter, and
$$\sqrt{C^2 - V^2\sin^2\theta} + V\cos\theta \qquad (15)$$

---

[4] Generally, given that the refractive coefficient of air is very close to 1, the atmosphere is treated as empty when it comes to measuring the speed of light in a medium of high refractive coefficient. But, in absolute terms, the air is also a refractive medium. An experiment of Hoek's type may be considered, in which, one part of the circuit would consist of vacuum, and the other part of air, considered refractive medium. *n* being the refractive coefficient of air, our analysis shows that the speed of light in it measured on Earth would yield $\frac{C}{n}\sqrt{1-\frac{V^2}{C^2}(1-\frac{1}{n^2})} - \frac{V}{n^2}$ in the direction of the Earth absolute velocity and $\frac{C}{n}\sqrt{1-\frac{V^2}{C^2}(1-\frac{1}{n^2})} + \frac{V}{n^2}$ in the opposite direction, like any other refractive medium..

when it points in the opposite direction [6, 14, 34].

From the expressions (12) and (13) and (14) and (15), the equality of the times observed for the two light paths, whatever the orientation of the device in Hoek's experiment, requires that:

$$\frac{L}{\sqrt{C^2 - V^2 \sin^2 \theta} - V \cos \theta} + \frac{L}{\frac{C}{n} + \frac{V}{n^2} \cos \theta + \varepsilon}$$
$$-(\frac{L}{\sqrt{C^2 - V^2 \sin^2 \theta} + V \cos \theta} + \frac{L}{\frac{C}{n} - \frac{V}{n^2} \cos \theta + \varepsilon}) = 0 \qquad (16)$$

where $\varepsilon$ denotes the terms to be added to Hoek's formula so that it amounts to second order.

Denoting $\sqrt{C^2 - V^2 \sin^2 \theta} = A$, expression (16) gives successively:

$$\frac{2V \cos \theta}{A^2 - V^2 \cos^2 \theta} - \frac{\frac{2V}{n^2} \cos \theta}{(\frac{C}{n} + \varepsilon)^2 - \frac{V^2}{n^4} \cos^2 \theta} = 0$$

$$A^2 - V^2 \cos^2 \theta - n^2 (\frac{C}{n} + \varepsilon)^2 + \frac{V^2}{n^2} \cos^2 \theta = 0$$

$$n^2 \varepsilon^2 + 2Cn\varepsilon + V^2 - \frac{V^2}{n^2} \cos^2 \theta = 0.$$

Resolving the second degree equation yields:

$$\varepsilon = \frac{-2Cn \pm \sqrt{4C^2 n^2 - 4n^2 V^2 (1 - \frac{1}{n^2} \cos^2 \theta)}}{2n^2}.$$

Retaining only the plus sign which has only a physical meaning, this expression simplifies to:

$$\varepsilon = -\frac{C}{n} + \frac{C}{n} \sqrt{1 - \frac{V^2}{C^2}(1 - \frac{1}{n^2} \cos^2 \theta)},$$

so that the modified expressions of Hoek's formulas which take into account the second order terms are:

$$C_1 = -\frac{V}{n^2} \cos \theta + \frac{C}{n} \sqrt{1 - \frac{V^2}{C^2}(1 - \frac{1}{n^2} \cos^2 \theta)} \qquad (17)$$

when the projection of the velocity vector of the light signal on the Earth absolute velocity points in the same direction as the latter, and

$$C_2 = +\frac{V}{n^2} \cos \theta + \frac{C}{n} \sqrt{1 - \frac{V^2}{C^2}(1 - \frac{1}{n^2} \cos^2 \theta)} \qquad (18)$$

when it points in the opposite direction..

These expressions can also be written as:

$$C_1 = \frac{C}{n} - \frac{V}{n^2} \cos \theta - \frac{V^2}{2Cn}(1 - \frac{1}{n^2} \cos^2 \theta)$$

$$C_2 = \frac{C}{n} + \frac{V}{n^2}\cos\theta - \frac{V^2}{2Cn}(1 - \frac{1}{n^2}\cos^2\theta).$$

Consistency of the derivation

1. $C_1$ and $C_2$ must obey the following conditions [6, 14, 34]:

When the refractive coefficient is equal to 1, $C_1$ must reduce to

$$\sqrt{C^2 - V^2 \sin^2\theta} - V\cos\theta$$

and $C_2$ must reduce to

$$\sqrt{C^2 - V^2 \sin^2\theta} + V\cos\theta.$$

2. The speed of light must tend toward $C/n$, whatever the direction, when the mathematical expression of the absolute speed of the platform ($V$) which supports the refractive medium tends toward zero.

3. When $\theta = 0$ and $n=1$, the speed of light must reduce to
$$C - V$$
when the light signal travels in the same direction as the Earth absolute velocity, and to
$$C + V.$$
when it travels in the opposite direction.

4. When $n=1$ and the light velocity vector points in a direction $\perp$ to the Earth absolute velocity, the speed of light in this direction must be equal to $\sqrt{C^2 - V^2}$.

5. In the oblique path covered by the light signal in the aether, in Michelson-Morley experiment (Figure1), the speed of light which differs from $C/n$ when $V$ is not null, should tend toward $C/n$ if the mathematical expression of $V$ was tending toward zero (see below).

We can easily verify that the expressions $C_1$ and $C_2$, in (17) and (18) obey the conditions 1 to 4.

For the condition 5:

We can check that for $\theta = \pi/2$, expressions (17) and (18) are equal to $\frac{C}{n}\sqrt{1 - \frac{V^2}{C^2}}$. The speed of light in the oblique path covered by the light ray in the aether frame in Michelson-Morley experiment has for square:

$$C'^2 = \frac{C^2}{n^2}(1 - \frac{V^2}{C^2}) + V^2,$$

thus:

$$C' = \sqrt{\frac{C^2}{n^2} + V^2(1 - \frac{1}{n^2})}.$$

If the mathematical expression of $V$ tends toward zero, this expression reduces to $C/n$. Therefore, the expressions (17) and (18) obey the condition 5.

**Appendix 2**
**Further objections against the aether drag hypothesis.**

The concept of an aether entrained by the motion of celestial bodies is still alive among some groups of physicists despite the arguments which have been given against its existence.

To maintain their belief, the supporters of the entrainment don't hesitate to cast doubt to the theories and experiments which support the inverse thesis or to their interpretation (Lorentz aether theory and experiments performed by Bradley, Lodge, Hammar, Herrmann, Demjanov, Cahill among others).

In order to settle the issue, it is worth analysing the consequences of such a hypothetical entrainment. To this end, let us suppose that the Earth (and the other celestial bodies) are surrounded by a crown of aether attracted by gravitation. Of course such a hypothesis implies that the aether possesses mass. However, at a distance from the celestial bodies, the aether can no longer be subjected to such attraction and an aether wind blows to which the planets are confronted. This is what the reasoning shows and what the experiments of Smoot have verified [33].

With respect to the source of radiation Smoot and Gorenstein's experiments have found that the solar system moves at a velocity close to 370 km/sec, (namely 1,335,600 Km/h) which corresponds to an aether wind of this magnitude [33].

In the course of its travel, the front of the Earth with its hypothetic crown of attracted aether would be faced to the pressure exerted by the non-entrained aether which blows at more than 1 million Km/h. In all likelihood, across from this wind, the crown of aether could not be static, it would be pushed in the direction opposite to the Earth absolute velocity in such a way that the entrainment would be thwarted. So, even though one gives credit to an aether subject to gravitation, the entrainment is confronted to significant obstacles.

Note that a massive aether subject to gravitation would certainly have also exerted a pressure on the air molecules giving rise to an atmospheric wind, which ultimately could have expelled the atmosphere. The absence of such an effect argues in favour of a non massive aether, whose action on matter at low absolute speeds is negligible, strengthening the point of view developed by Lorentz.